\documentclass[aps,prl,twocolumn, superscriptaddress,floatfix,nofootinbib,longbibliography]{revtex4-2}
\usepackage{graphics}
\usepackage{epsfig}
\usepackage{times}
\usepackage{bm}
\usepackage{bbm}
\usepackage{braket}
\usepackage{color,float}

\newcommand{\zero}{\scalebox{0.8}[1.5]{\kern.1mm$\mathbbm{o}$}}

\usepackage{amsmath}	

\begin{document}


\title{Emergence of vortex state in the $S=1$ Kitaev-Heisenberg model with single-ion anisotropy}
	
\author{Ayushi Singhania} 
\affiliation{Institute for Theoretical Solid State Physics, IFW Dresden, 01069 Dresden, Germany}

\author{Jeroen van den Brink} 
\affiliation{Institute for Theoretical Solid State Physics, IFW Dresden, 01069 Dresden, Germany}
\affiliation{Department of Physics, Technical University Dresden, 01069 Dresden, Germany}

\author{Satoshi Nishimoto} 
\affiliation{Institute for Theoretical Solid State Physics, IFW Dresden, 01069 Dresden, Germany}
\affiliation{Department of Physics, Technical University Dresden, 01069 Dresden, Germany}

\begin{abstract}
The search for Kitaev spin liquid states has recently broadened to include a number of honeycomb materials with integer spin moments. The qualitative difference with their spin-1/2 counterparts is the presence of single-ion anisotropy (SIA). This motivates our investigation of  the effects of SIA on the ground state of the spin-$1$ Kitaev-Heisenberg (KH) model using the density-matrix renormalization group which allows construction of detailed phase diagrams around the Kitaev points. We demonstrate that positive out-of-plane SIA induces an in-plane vortex state without the need for off-diagonal interactions. Conversely, negative SIA facilitates the emergence of a {\it ferromagnetic} state in presence of antiferromagnetic Heisenberg interactions, while a {\it N\'eel} state can emerge for ferromagnetic Heisenberg coupling. These findings, pertinent even for weak SIA, not only enhance our theoretical understanding of the spin-1 KH model but also propose experimental prospects for observing these novel magnetic states in material realizations.
\end{abstract}

\maketitle
\textit{Introduction}---
The Kitaev model, initially formulated for spin-$1/2$ particles on a honeycomb lattice, offers a unique framework conducive to theoretical exploration due to its exact solvability~\cite{Kitaev2006}. Its ground state, known as the Kitaev spin liquid (KSL), exhibits fractionalized excitations such as Majorana fermions. The realization of Kitaev-type interactions in $d^5$ transition metal compounds with strong spin-orbit coupling (SOC) paved the way for a burgeoning field of research on Kitaev materials~\cite{Jackeli2009, Trebst2022}. The extension to the Kitaev-Heisenberg (KH) model, which incorporates conventional Heisenberg interactions, enables a more accurate portrayal of magnetic behaviors in actual compounds~\cite{Winter2017}.
Recent experimental advancements have broadened the search beyond spin-$1/2$ to spin-$1$ systems, unearthing potential Kitaev physics in novel materials such as Na$_2$Ni$_2$TeO$_6$~\cite{stavropoulos2019microscopic, Samarakoon2021, bera2022magnetism}, A$_3$Ni$_2$SbO$_6$ (A=Li, Na)~\cite{zvereva2015zigzag}, and KNiAsO$_4$  ~\cite{taddei2023zigzag}. The complexity of the systems for $S>1/2$ is heightened due to their lack of an exact solution and the absence of a Majorana fermion description. Instead, several numerical studies have been undertaken to elucidate the fundamental properties of the spin-$1$ Kitaev model~\cite{koga2018ground,oitmaa2018incipient,dong2020spin,Zhu2020,Hickey2020,Lee2020,Khait2021,bradley2022instabilities,fukui2022ground,pohle2023spin, consoli2020heisenberg, khait2021characterizing,zhu2020magnetic}. Notably, the broader KSL region found in the spin-$1/2$ system~\cite{Chaloupka2010} is constricted in spin-$1$ systems, with the latter exhibiting enhanced stability for magnetically ordered phases due to reduced quantum fluctuations ~\cite{fukui2022ground}. At the Kitaev limit, the existence of a generalized hexagonal plaquette operator preserves the $Z_2$ gauge structure, leading to vanishing spin-spin correlations beyond nearest neighbors as in the spin-$1/2$ model~\cite{baskaran2007exact,baskaran2008spin}.

A pivotal aspect of quantum spin systems with $S>1/2$ is the presence of single-ion anisotropy (SIA), which may critically affect magnetic characters ~\cite{stavropoulos2021magnetic, jin2022unveiling,chen2023triple, luo2023spontaneous, xu2018interplay}. The theoretical underpinning of this term is captured by the integration of a SOC term into the Hamiltonian, which subsequently induces SIA. Particularly in Kitaev materials, pronounced SOC may contribute to significant SIA. Although the implications of SIA have been probed in the context of the Kitaev limit~\cite{bradley2022instabilities, jin2022unveiling}, the consequences integrating SIA into the spin-$1$ KH model, as pertinent to material realizations, has remained to be comprehensively explored.

In this Letter, we study the ground state properties of the spin-$1$ KH model on a honeycomb lattice with out-of-plane SIA using the density-matrix renormalization group (DMRG). By analyzing the energy spectrum, spin structure factor, chiral vector, and flux expectations, we present detailed phase diagrams surrounding KSL phases, which reveal the intricate interplay between the Kitaev, Heisenberg, and SIA terms. Our study leads to two significant insights: the induction of an in-plane aligned vortex state by positive SIA, even in the absence of off-diagonal interactions, and the surprising stabilization of a ferromagnetic (FM) state under antiferromagnetic (AFM) Heisenberg coupling for negative SIA and vice versa. These novel magnetic phenomena can manifest even with relatively small SIA, thus presenting a viable avenue for experimental observation in real materials. 

\begin{figure}[tbh]
	\centering
	\includegraphics[width=1.0\columnwidth]{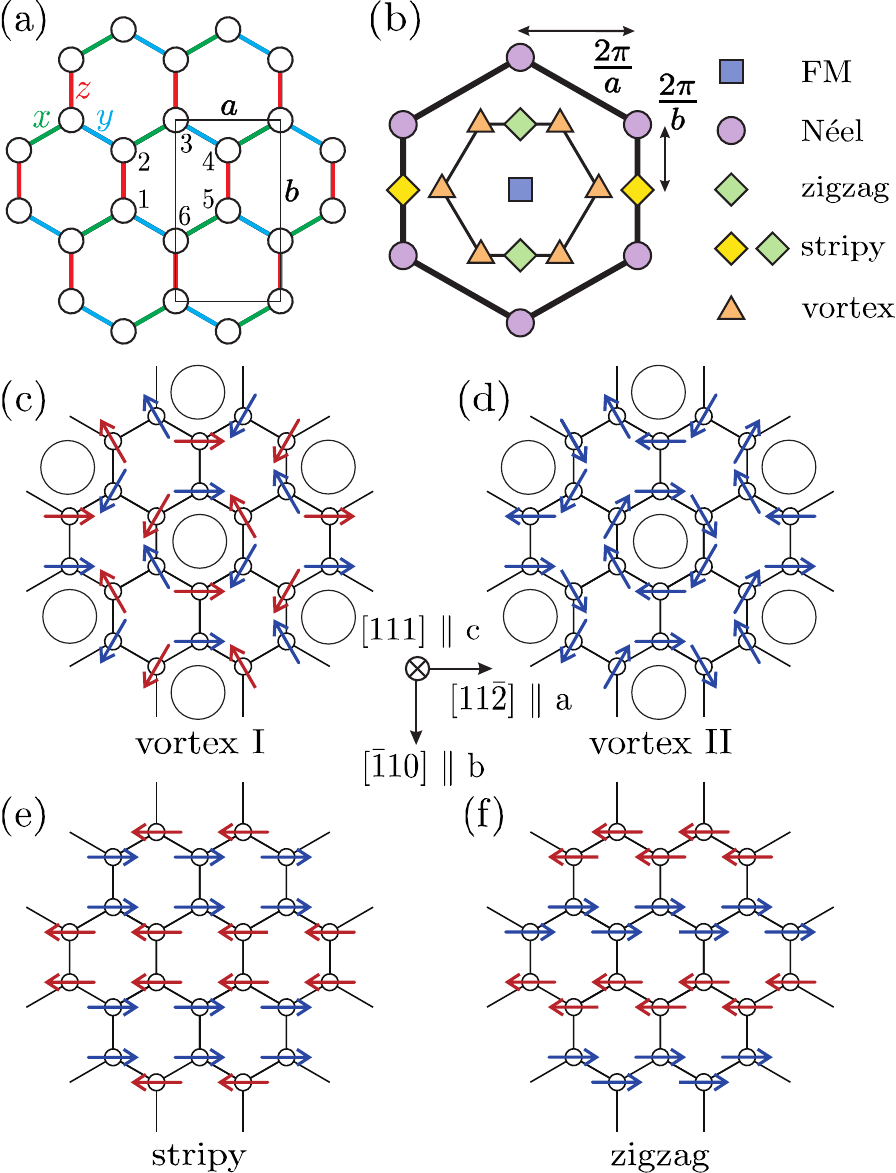} 
	\caption{(a) $C_3$-symmetric 24-site OBC cluster used in our DMRG calculations, where the thin-lined rectangle denotes the structural unit cell. The numbers 1-6  labeling for the plaquette operator and chiral vector. (b) Reciprocal space diagram showing Bragg peak positions for various magnetic phases (inner hexagon shows ﬁrst Brillouin zone of the honeycomb lattice). (c) Schematic structure of the vortex state. 
 }
	\label{fig:lattice}
\end{figure}

\textit{Model}---
We consider a spin-$1$ KH model on a honeycomb lattice. The Hamiltonian is expressed as

\begin{align}
\hspace{-0.4cm}
\mathcal{H}_{\rm KH}=2K\sum_{\langle ij \rangle\gamma} S_i^\gamma S_j^\gamma + J \sum_{\langle i,j \rangle} \mathbf{S}_i \cdot \mathbf{S}_j
+ D \sum_{i} (\mathbf{n} \cdot \mathbf{S}_i)^2
\hspace{-0.2cm}
\label{eq:KH_Ham}
\end{align}
where $S_i^\gamma$ is the $\gamma$ ($=x$, $y$, or $z$) component of the spin-$1$ operator $\mathbf{S}_i$ at site $i$, $\mathbf{n}$ represents the direction of SIA and the strength of SIA is controlled by $D$, and $J$ and $K$ are the Heisenberg and Kitaev interactions, respectively. For convenience, we introduce an angle parameter $\theta \in[0, 2\pi]$, setting $J=\cos \theta$ and $K=\sin \theta$. Focusing on the most generally plausible scenario, we consider the case where $\mathbf{n}$ is perpendicular to the honeycomb plane, i.e., $\mathbf{n} \parallel [111]$. This type of SIA is naturally anticipated from trigonal distortions ~\cite{stavropoulos2021magnetic, liu2022theoretical}. On the other hand, the microscopic mechanism identified for higher-$S$ Kitaev candidate materials suggests the absence of off-diagonal $\Gamma$ interactions~\cite{stavropoulos2019microscopic}. Thus, our study focuses on the spin-$1$ KH model with SIA. 
It is noteworthy that a transformation of Hamiltonian \eqref{eq:KH_Ham} from $xyz$ to $abc$ coordinates results in the emergence of apparent off-diagonal terms (see the Supplementary Material). This could be pertinent to the emergence of vortex states discussed below.

\textit{Method}---
We perform DMRG calculations on a $C_6$-symmetric, i.e., spatially-anisotropic, cluster with open boundary conditions (OBC). The use of OBC allows for calculations of high precision and avoids the finite-size effects that typically skew the interpretation of anisotropy-sensitive phases such as zigzag or stripy patterns, as noted by recent work~\cite{kadosawa2023}. For a broad survey of the phase space and to obtain a comprehensive phase diagram, we opt for a 24-site cluster configuration depicted in Fig.~\ref{fig:lattice}(a). Although a finite-size scaling is necessary to confirm long-range ordering conclusively, the intrinsic competition between the local terms as found in Hamiltonian \eqref{eq:KH_Ham} can still be examined using a cluster of this size. The maximum bond dimension is set to $1000$, so that the maximum truncation error is less than $\sim 1 \times 10^{-6}$. We have verified consistency with prior studies at $D=0$~\cite{dong2020spin,fukui2022ground} and $J=0$~\cite{bradley2022instabilities} (see the Supplementary Material). Subsequently, in order to confirm the reliability of our method, our results are compared to those obtained using larger 37-site and 54-site clusters for representative parameters of each phase. Supplementary data supporting the computational veracity, including convergence tests and additional DMRG outputs, are provided in the Supplementary Material. Our results were obtained with ITensor package~\cite{itensor}.


\textit{Physical quantities}---
To delineate the ground-state phase diagram, we compute the second derivative of the ground-state energy $E$ with respect to the control parameters $\lambda$ (which can be $\theta$ or $D$):
\begin{align}
	E''(\lambda)=\frac{\partial^2 E}{\partial \lambda^2},
	\label{eq:d2E}
\end{align}
and identify phase transitions where $E''(\lambda)$ peaks. Additionally, we evaluate the expectation value of the plaquette operator $W_{\rm p}$ given by:
\begin{align}
	W_{\rm p} = \exp\left[i\pi (S^x_1 + S^y_2 + S^z_3 + S^x_4 + S^y_5 + S^z_6)\right],
	\label{eq:Wp}
\end{align}
which serves as an order parameter distinguishing the topological phases, with its eigenvalues deviating from $\pm 1$ indicating transition from the KSL regime. The chiral vector $\boldsymbol{v}_{\rm c}$, which probes spin chirality, is calculated as:
\begin{align}
	\mbox{\boldmath $v$}_{\rm c} =  \sum_{i=1}^3 \mbox{\boldmath $S$}_{2i-1} \times \mbox{\boldmath $S$}_{2i+1}=\sum_{i=1}^3 \mbox{\boldmath $S$}_{2i} \times \mbox{\boldmath $S$}_{2i+2},
	\label{eq:vc}
\end{align}
where the labeling of $i$ is shown in Fig.~\ref{fig:lattice}(a). $\boldsymbol{v}_{\rm c}$ provides insight into 120$^\circ$ magnetic order and/or vortex state. We show average taken over all plaquettes for $| \mathbf{n}  \cdot \boldsymbol{v}_{\rm c} |$. We further quantify the static spin structure factor $S(\boldsymbol{Q})$, which reflects spin correlations at wave vector $\boldsymbol{Q}$ across the cluster:
\begin{align}
	S(\boldsymbol{Q})=\frac{1}{N}\sum_{ij} \langle \boldsymbol{S}_i \cdot \boldsymbol{S}_j \rangle e^{i \boldsymbol{Q} \cdot (\mathbf{r}_i-\mathbf{r}_j)},
	\label{eq:SQ}
\end{align}
where $N$ is the total number of lattice sites and $\mathbf{r}_i$ denotes the position of site $i$. A peak in $E''(\lambda)$ typically indicates a phase transition, and by examining the behavior of $S(\boldsymbol{Q})$, we can determine the nature of spin ordering within the various phases. The deviation of $W_{\rm p}$ from $\pm 1$ provides a quantitative measure of the robustness of the KSL phase.

\begin{figure}[tbh]
	\centering
	\includegraphics[width=1.0\columnwidth]{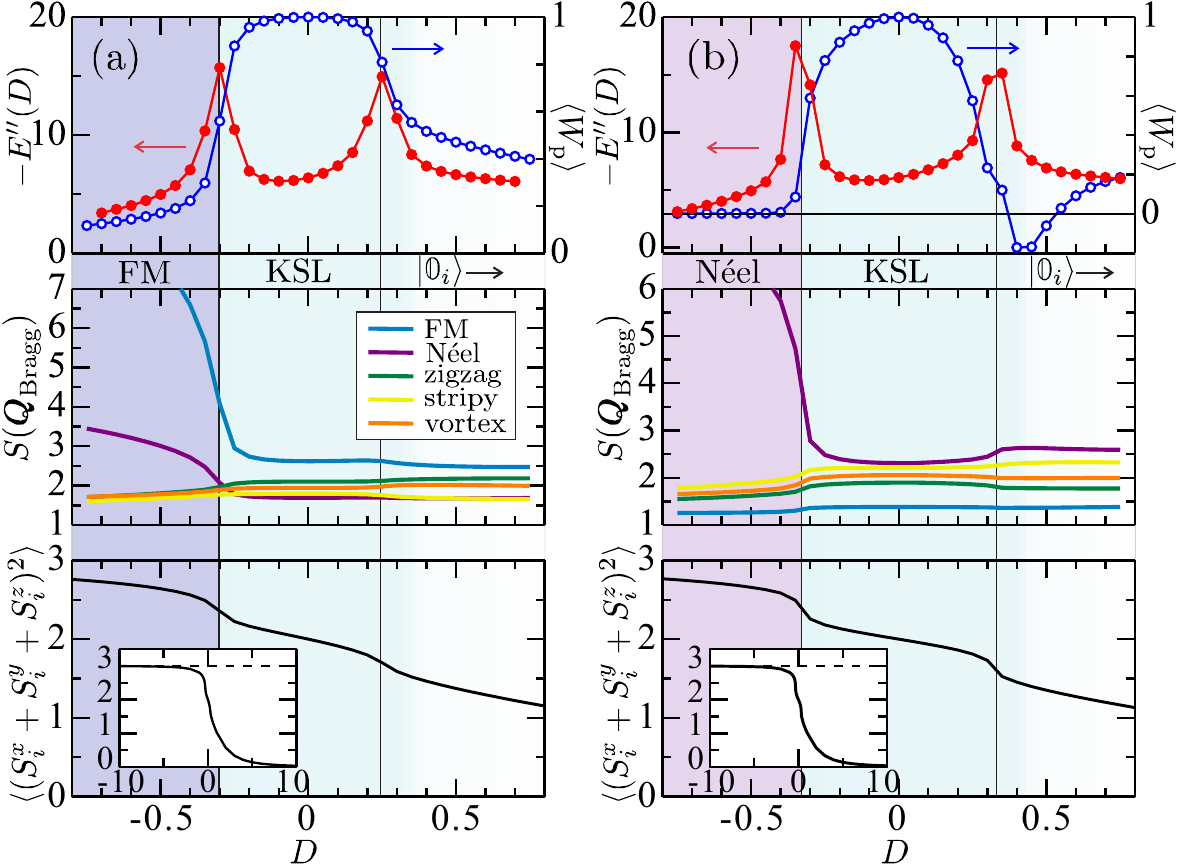} 
	\caption{DMRG results for the second derivative of ground-state energy, expectation value of the flux operator (top), intensity of the spin structure factor at representative Bragg peak positions (middle), and expectation value of $(S^x_i+S^y_i+S^z_i)^2$ (bottom) as a function of $D$ with fixed (a) $\theta=1.5\pi$ and (b) $\theta=0.5\pi$. Inset of the bottom figure shows $\langle (S^x_i+S^y_i+S^z_i)^2 \rangle$ for extended range of $D$.
	}
	\label{fig:vary_D}
\end{figure}

\textit{Reduction of spin-1 degrees of freedom in large $|D|$ regions}---
In the large $D$ limit, the ground state can be expressed as a product over site-specific states, $|\psi_{\rm gs}\rangle=\prod_i|\zero_i\rangle$ with
\begin{align}
	|\zero_i\rangle=\frac{1}{\sqrt{3}}(|x\rangle+|y\rangle+|z\rangle),
	\label{eq:zero}
\end{align}
where $|x\rangle=-\frac{1}{\sqrt{2}}(|1\rangle-|-1\rangle)$, $|y\rangle=\frac{i}{\sqrt{2}}(|1\rangle+|-1\rangle)$, and $|z\rangle=|0\rangle)$~\cite{koga2018ground,bradley2022instabilities}. Here, $|1\rangle$, $|0\rangle$, and $|-1\rangle$ represent the eigenstates of the z-component of spin-1 particle. The expectation values for spin components and spin-spin correlations in $|\zero_i\rangle$ vanish, i.e., $\langle \zero_i|S_i^\alpha|\zero_i\rangle = 0$ and $\langle\psi_{\rm gs}|S_i^\alpha S_j^\alpha|\psi_{\rm gs}\rangle=0$ for any sites $i$, $j$ and $\alpha = x, y, z$. Thus, the system is in a disordered state at large $D$ regions.

For substantially negative values of $D$, the effective spin-1 degrees of freedom resemble to that of Ising model, due to the preferential alignment of spins along the [111] direction or its antipodal direction [$\bar{1}\bar{1}\bar{1}$]. The pertinent spin states can be defined in this limit as:
\begin{align}
	&|\Uparrow\rangle=\frac{1}{\sqrt{3}}(e^{i\frac{2}{3}\pi}|x\rangle+e^{i\frac{4}{3}\pi}|y\rangle+|z\rangle)
	\label{eq:Isinga}\\
	&|\Downarrow\rangle=\frac{1}{\sqrt{3}}(e^{i\frac{4}{3}\pi}|x\rangle+e^{i\frac{2}{3}\pi}|y\rangle+|z\rangle).
	\label{eq:Isingb}
\end{align}
This yields the expectation values of $\langle\Uparrow|S_i^\gamma|\Uparrow\rangle = \frac{1}{\sqrt{3}}$ and $\langle\Downarrow|S_i^\gamma|\Downarrow\rangle = -\frac{1}{\sqrt{3}}$ for $\gamma = x, y, z$ (see the Supplemental Material for details).

\textit{The $J=0$ limit}---
We commence our analysis by considering the limit of vanishing Heisenberg coupling $J$. The system manifests the KSL ground state for $D=0$. The phase evolution with varying $D$ for a fixed $\theta=1.5\pi$ (corresponding to $J=0$, $K=-1$) is depicted in Fig.~\ref{fig:vary_D}(a). We find that $E''(D)$ has two peaks at $D\approx-0.33$ and $0.32$. Within the interval $-0.33 \lesssim D \lesssim 0.32$, the system retains the KSL state, corroborated by the persistent topological invariant $\langle W_{\rm p} \rangle \sim 1$. When $D$ falls below $-0.33$, $\langle W_{\rm p} \rangle$ sharply diminishes close to zero with a steep increase of the FM Bragg peak intensity, signifying a phase transition to an FM ordered state. This occurrence of long-range order, induced by negative SIA, is a particularly remarkable consequence of effective Ising spins $|\Uparrow\rangle$ and $\Downarrow\rangle$. When subject to Kitaev interactions, negative and positive values of $K$ foster FM and N\'eel orders, respectively. For both cases, the ordering effectuates an energy reduction by $|K|$, setting the ground-state energy per spin at $\varepsilon = D - |K|$. This clearly explains why the addition of a $D$ term to the Kitaev model is enough to create a long-range order~\cite{bradley2022instabilities}.

In contrast, for $D\gtrsim0.32$, despite a marked decrease of $\langle W_{\rm p} \rangle$ to approximately $0.6$, no discernible rise in Bragg peak intensity occurs. The systematic decline of $\langle W_{\rm p} \rangle$ upon increasing $D$ supports the interpretation of this regime as a crossover from the KSL phase to a potential disordered phase. Indeed, in the vicinity just beyond $D\approx0.32$, the persistence of strong nearest-neighbor correlations still affirms the KSL characteristics, as detailed in the Supplementary Material. 

The strong contrast between the abrupt phase change at $D \approx -0.33$ and the more gradual crossover beyond $D \approx 0.32$ is clarified upon examining the squared [111] spin component, $(S_i^x+S_i^y+S_i^z)^2$. This quantity achieves a value of $3$ when spin-1 is fully collapsed into the Ising degrees of freedom, and it becomes $0$ when the system is in a complete zero state. As shown in the bottom figure of Fig.~\ref{fig:vary_D}(a), this value is already close to $3$ at $D \approx -0.33$, and it rather gradually approaches $0$ as $D$ exceeds $0.33$.

Figure~\ref{fig:vary_D}(b) presents analogous findings for $\theta = 0.5\pi$. The FM order is now replaced by AFM N\'eel order as $D$ turns negative, yet the resemblance to the phase behavior for $\theta = 1.5\pi$ is striking, demonstrating the robustness of these phenomena across the chosen parametric spectrum.

\begin{figure}[tbh]
	\centering
	\includegraphics[width=1.0\columnwidth]{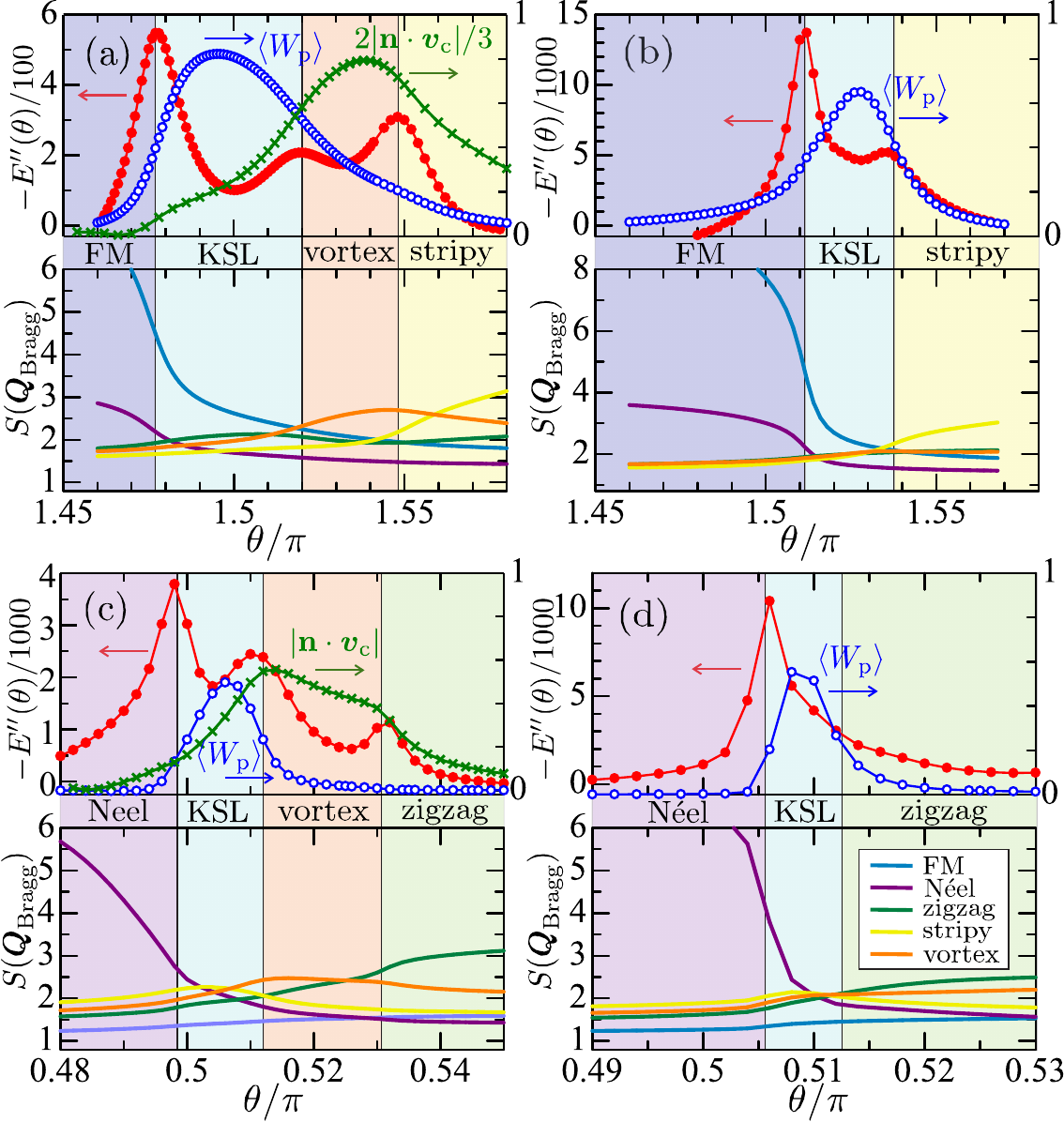} 
	\caption{DMRG results for the second derivative of ground-state energy, expectation value of the flux operator, the magnitude of chiral vector (top), intensity of the spin structure factor at representative Bragg peak positions (bottom) around (a,b) FM and (c,d) AFM Kitaev points. The values of $D$ are fixed at (a) $0.25$, (b) $-0.5$, (c) $0.3$, and (d) $-0.5$.}
	\label{fig:vary_theta}
\end{figure}

\begin{figure}[tbh!]
	\centering
	\includegraphics[width=0.6\columnwidth]{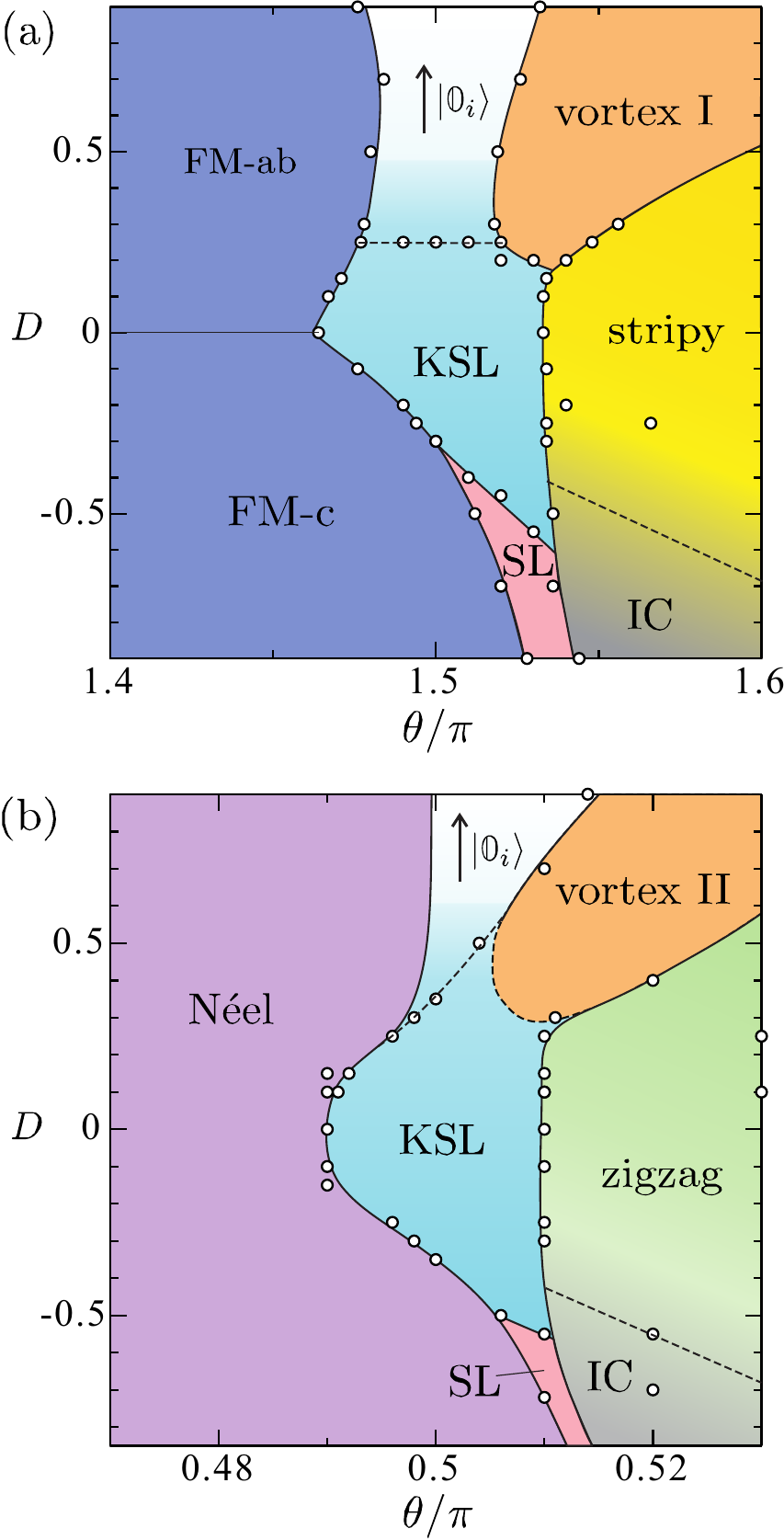} 
	\caption{Ground-state phase diagram of the system \eqref{eq:KH_Ham} in the $D$$-$$\theta$ space around (a) FM and (b) AFM Kitaev points. Solid lines indicate phase boundary and dotted lines denote more crossover-like transition. The circles are obtained from the peak positions in the second derivative of ground-state energy.
	}
	\label{fig:PD}
\end{figure}

\textit{Phase diagram in the $D$$-$$\theta$ space}---
To determine the phase distribution near the FM and AFM Kitaev points, we calculate $E''(\lambda)$, $\langle W_{\rm p} \rangle$, $\mbox{\boldmath $v$}_{\rm c}$, and $S(\mbox{\boldmath $Q$})$ over the $D$ and $\theta$ parameter space. At $D=0$, phase transition sequences near the AFM Kitaev point—specifically, N\'eel ($\theta/\pi \lesssim 0.491$) to AFM KSL ($0.491 \lesssim \theta/\pi \lesssim 0.509$) to zigzag ($\theta/\pi \gtrsim 0.509$)—are established. Corresponding transitions near the FM Kitaev point delineate FM ($\theta/\pi \lesssim 1.464$) to FM KSL ($1.464 \lesssim \theta/\pi \lesssim 1.533$) to zigzag ($\theta/\pi \gtrsim 1.533$). The narrowed KSL regions, as compared to those in the spin-$1/2$ case, reflect reduced quantum fluctuations and are in line with literature~\cite{fukui2022ground}. To explore the effects of non-zero $D$ values, we plot $E''(\theta)$, $\langle W_{\rm p} \rangle$, $|\mathbf{n} \cdot \mbox{\boldmath $v$}_{\rm c}|$, $S(\mbox{\boldmath $Q$})$ around the FM Kitaev point with fixed $D=0.25$ and $-0.5$ in Fig.~\ref{fig:vary_theta}(a,b), as examples. Each transition point is clearly indicated by peaks in $E''(\theta)$, with corresponding changes of peak positions or amplitudes in $S(\mbox{\boldmath $Q$})$. The composite ground-state phase diagram in the $D$$-$$\theta$ space is depicted in Fig.~\ref{fig:PD}(a), highlighting two prominent observations.

First, a vortex phase emerges uniquely between the zero and stripy phases, exhibiting a pronounced enhancement of the chiral vector [$|\mathbf{n} \cdot \boldsymbol{v}_{\rm c}| = 3\sqrt{3}/2 \approx 2.60$ for the idealized vortex configuration as seen in Fig.~\ref{fig:lattice}(c)]. While this state is essentially the same as observed in the spin-$1/2$ honeycomb-lattice $J$-$K$-$\Gamma$-$\Gamma'$ model~\cite{Chaloupka2015}, it is surprising that such a vortex state stabilizes despite the absence of off-diagonal interactions. In the phase diagram, one can discern that the region of the vortex phase is primarily confined to values of $D$ greater than that of the stripe phase. This can be qualitatively explained using a classical model: as $D$ becomes increasingly positive, the orientation of the spins tends to be constrained parallel to the ab-plane. Assuming that all spins lie within the ab-plane, the energies per site of the vortex-I state [Fig.~\ref{fig:lattice}(c)] and the stripy state [Fig.~\ref{fig:lattice}(e)] are $E^{\rm cl}({\rm vortex\textrm{-}I)}=K/2$ and $E^{\rm cl}({\rm stripy})=J/2+K/6$, respectively. This means that the energy gain from the Kitaev term is maximized by forming a vortex in a hexagon. Thus, as $D$ increases, a phase transition from the stripy to the vortex phases at a certain point seems likely for $-K>>J>0$. This phenomenon may be related to an in-plane vortices with easy-plane anisotropy~\cite{Wysin1994}. As $D$ continues to increase beyond this, the system transitions to a zero phase due to the absence of spin correlations.

Second, we observe that when $D$ is negative, the FM phase expands anomalously into the AFM $J$ parameter space, i.e., $\theta>1.5\pi$, a phenomenon that is not seen in the simple KH model. This can be explained by the reduction of degrees of freedom from spin-1 to Ising. As mentioned above, when the spin basis has only either state with Eq.~\eqref{eq:Isinga} or with Eq.~\eqref{eq:Isingb}, the Kitaev term can lower the energy by $|K|$ with forming an FM state. If an AFM $J$ is introduced to this FM state, the energy increase can be limited to $3J/2$. Thus, as long as $|K|>3J/2$, the FM phase can maintain its stability. The SL phase inbetween FM and IC (or stripy) phases may be a consequence of the competition of FM Kitaev and AFM Heisenberg interactions. In the large negative $D$ limit, the phase boundary between FM and SL phases approaches $\theta \approx 1.81\pi$. For more details, please refer to the Supplementary Material.

Turning to the vicinity of the AFM Kitaev point, the structure of phase diagram, as depicted in Fig.~\ref{fig:vary_theta}(c,d) and Fig.~\ref{fig:PD}(b), nearly mirrors the one around the FM Kitaev point with the N\'eel and zigzag phases supplanting the FM and stripy phases, respectively. The vortex-II state, as illustrated in Fig.~\ref{fig:lattice}(d), exhibits a spin structure distinct from the vortex-I state due to the spin orientations on the inter-vortex bond being contingent upon the sign of $J$. A thorough explanation of distinguishing vortex-II from vortex-I states is provided in the Supplementary Material. Assuming a restriction of spin alignment within the ab-plane, the per-site energies of the vortex-II state [Fig.~\ref{fig:lattice}(d)] and the zigzag state [Fig.~\ref{fig:lattice}(f)] are calculated to be $E^{\rm cl}({\rm vortex\textrm{-}II})=-K/2$ and $E^{\rm cl}({\rm zigzag})=-J/2-K/6$, respectively. For a sufficiently positive $D$, the vortex-II state is anticipated to supersede the stability of the zigzag state for $K \gg -J > 0$. Moreover, at negative $D$, the unusual presence of the N\'eel phase within the region of FM Heisenberg interactions is rationalized by the dominance of the AFM Kitaev interaction benefits. Namely,  the energy gain from the N\'eel configuration, attributed to positive $K$ values, compensates for the energy loss introduced by an FM character of $J$.

\textit{Summary}---
Using DMRG we have considered the spin-$1$ honeycomb-lattice KH model with SIA $D$ and investigated its ground states via the second derivative of ground-state energy, chiral vector, spin structure factor, and the expectation values of the flux operator and $(S^x_i+S^y_i+S^z_i)^2$, uncovering rich phase diagrams that extend the boundaries of known Kitaev physics.
Our results underscore the pronounced impact of minimal SIA on the ground state proximate to both FM and AFM Kitaev points, delineating novel ground-state domains. A vortex state emerges as a stable configuration when shifting from conventional zigzag or stripe order towards a singular zero state, facilitated by positive $D$ favoring in-plane spin orientations. Remarkably, for $D<0$, we report the stabilization of an FM phase amidst AFM Heisenberg interactions and vice versa, challenging conventional expectations. This is attributable to negative $D$ imposing spin orientation constraints, allowing Kitaev interactions to promote magnetic order over SL states -- a phenomenon absent in the canonical KH model.

The discovery of these magnetic orders, induced by a slight SIA, may pave the way for further experimental pursuits and material-specific calculations. Indeed, it has been reported that SIA on the order of $\sim 1$ meV is plausible in Na$_2$Ni$_2$TeO$_6$~\cite{Samarakoon2021}. Thus our findings carve out new pathways for the discovery and design of quantum magnets with exotic properties while enriching the theoretical landscape of high-spin Kitaev systems.

{\it Acknowledgements.---}
We thank Ulrike Nitzsche for technical assistance. This project is funded by IFW Excellence Programme 2020 and the German Research Foundation (DFG) via the projects A05 of the Collaborative Research Center SFB 1143 (project-id 247310070) and through the W\"urzburg-Dresden Cluster of Excellence on Complexity and Topology in Quantum Matter ct.qmat (EXC 2147, Project No. 39085490).

\end{document}


\title{Supplementary Material for\\ ``Emergence of vortex state in the $S=1$ Kitaev-Heisenberg model with single-ion anisotropy''}

\author{Ayushi Singhania$^1$, Jeroen van den Brink$^{1,2}$, and Satoshi Nishimoto$^{1,2}$}

\address{
$^1$ Institute for Theoretical Solid State Physics, IFW Dresden, 01069 Dresden, $^2$ Department of Physics, Technical University Dresden, 01069 Dresden, Germany
}

\date{\today}

\begin{abstract}
We here provide supplemental explanations and data on the following topics in relation to the main text: 1) The representation of eigenstates of a spin-1 site in the limits of $D=\infty$ and $D=-\infty$. 2) Transformation of the Hamiltonian from $xyz$ to $abc$ coordinates. 3) How the energy in the classical limit of each phase is expressed when the eigenstates obtained in 1) are assumed. 4) Observation of the Crossover from Kitaev spin liquid to the zero state. 5) A comparison with previous studies in the case of $D=0$. 6) Finite-size scaling of the ground-state energy at the Kitaev point. 7) Convergence of structure factor with bond dimension $\chi$. 8) A verification of the cluster size dependence of the spin structure factor at representative points of each phase. 9) An example analysis for identifying the vortex-I and vortex-II states.
\end{abstract}

\maketitle

\section{Splitting of spin-1 degrees of freedom due to single-ion anisotropy in [111]}

We consider how the spin-1 degrees of freedom are split due to single-ion anisotropy in [111]. The original spin-1 degrees of freedom for each site has three eigenstates $|1\rangle$, $|0\rangle$, $|-1\rangle$, and they are degenerate for an isolated site. When magnetic field $B$ and single-ion anisotropy $D$ parallel to the c-axis, i.e., [111], are applied on site $i$, the Hamiltonian is written as
\begin{align}
{\cal H}_i = \frac{D}{3}(S^x_i+S^y_i+S^z_i)^2-\frac{B}{\sqrt{3}}(S^x_i+S^y_i+S^z_i)
\label{ham}
\end{align}
where $S_i^\gamma$ is the $\gamma$ ($=x$, $y$, or $z$) component of the spin-$1$ operator $\mathbf{S}_i$ at site $i$. By diagonalizing this, we obtain three eigenstates:
\begin{align}
	|\, \zero\, \rangle = \frac{1}{\sqrt{3}}\left( - \frac{1-i}{\sqrt{2}}|1\rangle + |0\rangle + \frac{1+i}{\sqrt{2}}|-1\rangle\right)
	\label{zero}
\end{align}
with energy $\varepsilon=0$,
\begin{align}
	| \Uparrow \rangle = \frac{1}{\sqrt{3}}\left( \frac{\sqrt{3}+1}{2\sqrt{2}}(1-i)|1\rangle + |0\rangle + \frac{\sqrt{3}-1}{2\sqrt{2}}(1+i)|-1\rangle\right)
	\label{Up}
\end{align}
with energy $\varepsilon=D+B$, and
\begin{align}
	| \Downarrow \rangle = \frac{1}{\sqrt{3}}\left( \frac{-\sqrt{3}-1}{2\sqrt{2}}(1-i)|1\rangle + |0\rangle - \frac{\sqrt{3}-1}{2\sqrt{2}}(1+i)|-1\rangle\right)
	\label{Dn}
\end{align}
with energy $\varepsilon=D-B$. Assuming
\begin{align}
	| x \rangle &= -\frac{1}{\sqrt{2}}(|1\rangle-|-1\rangle)
	\label{x}\\
	| y \rangle &= \frac{i}{\sqrt{2}}(|1\rangle+|-1\rangle)
	\label{y}\\
	| z \rangle &= |0\rangle,
	\label{z}
\end{align}
Eqs.\eqref{zero}-\eqref{Dn} can be encapsulated as follows:
\begin{align}
	&|\, \zero\, \rangle=\frac{1}{\sqrt{3}}(|x\rangle+|y\rangle+|z\rangle)
	\label{eq:zero}\\
	&|\Uparrow\rangle=\frac{1}{\sqrt{3}}(e^{i\frac{2}{3}\pi}|x\rangle+e^{i\frac{4}{3}\pi}|y\rangle+|z\rangle)
	\label{eq:Isinga}\\
	&|\Downarrow\rangle=\frac{1}{\sqrt{3}}(e^{i\frac{4}{3}\pi}|x\rangle+e^{i\frac{2}{3}\pi}|y\rangle+|z\rangle).
	\label{eq:Isingb}
\end{align}
When $D<B$, the degrees of freedom of the spin-1 site drop to 2, as given by Eqs.\eqref{eq:Isinga} and \eqref{eq:Isingb}. In the large negative $D$ limit, the two degrees of freedom can be regarded as those for the Ising model. Introducing $B$, which commutes with single ion anisotropy, enables us to explicitly fix the Ising direction. Thus, the quantization axis is set to be parallel to the c-axis, i.e, perpendicular to the ab-plane. Thus, the up and down spins are given by $|\Uparrow\rangle$ and $|\Downarrow\rangle$, respectively, Accordingly, they lead to $\langle\Uparrow|S_i^x|\Uparrow\rangle=\langle\Uparrow|S_i^y|\Uparrow\rangle=\langle\Uparrow|S_i^z|\Uparrow\rangle=\frac{1}{\sqrt{3}}$ and
$\langle\Downarrow|S_i^x|\Downarrow\rangle=\langle\Downarrow|S_i^y|\Downarrow\rangle=\langle\Downarrow|S_i^z|\Downarrow\rangle=-\frac{1}{\sqrt{3}}$. Once the eigenstates \eqref{eq:zero}-\eqref{eq:Isingb} are obtained, we can set $B$ to zero in our discussion for the KH model.

Let us then briefly discuss what happens when the sites are connected by interactions. In the large $D$ limit, the ground state is unique as a simple product of Eq.~\eqref{eq:zero}, and the system is in a nonmagnetic ground state with all spin correlations and spn components to be zero. While for the large negative $D$ limit, 
the ground state is doubly degenerate independent of the N\'eel or FM order. In the case of FM order, the effect of single-ion anisotropy may be similar to that of an external magnetic field along [111]~\cite{khait2021characterizing,zhu2020magnetic}.

\section{Transformation of Hamiltonian from $xyz$ to $abc$ coordinates}

Since the direction of single-ion anisotropy is set to be along the c-axis ($\parallel [111]$), namely, perpendicular to the ab-plane, it is informative to provide our Hamiltonian in the $abc$ coordinate instead of $xyz$ one. As written in the main text, the original Hamiltonian for the $z$-bond is
\begin{align}
\mathcal{H}_{\rm KH}^z=J \sum_{\langle i,j \rangle} \mathbf{S}_i \cdot \mathbf{S}_j + 2K\sum_{\langle ij \rangle} S_i^z S_j^z + \frac{D}{3} \sum_{i} (S_i^x+S_i^y+S_i^z)^2.
\label{eq:KH_Ham_z}
\end{align}
Using the standard notations
\begin{align}
	S^a_i &= \frac{1}{\sqrt{6}}(S^x_i+S^y_i-2S^z_i)
	\label{Sa}\\
	S^b_i &= \frac{1}{\sqrt{2}}(-S^x_i+S^y_i)
	\label{Sb}\\
	S^c_i &= \frac{1}{\sqrt{3}}(S^x_i+S^y_i+S^z_i),
	\label{Sc}
\end{align}
we obtain
\begin{align}
	S^x_i &= \frac{1}{\sqrt{6}}(S^a_i-\sqrt{3}S^y_i+\sqrt{2}S^z_i)
	\label{Sx}\\
	S^y_i &= \frac{1}{\sqrt{6}}(S^a_i+\sqrt{3}S^y_i+\sqrt{2}S^z_i)
	\label{Sy}\\
	S^z_i &= \frac{1}{\sqrt{3}}(S^c_i-\sqrt{2}S^a_i).
	\label{Sz}
\end{align}
The Hamiltonian in the $abc$ coordinate system can be derived by substituting Eqs.~\eqref{Sx} through \eqref{Sz} into Eq.~\eqref{eq:KH_Ham_z}, as shown below.
\begin{align}
\mathcal{H}_{\rm KH}^z= J \sum_{\langle i,j \rangle} (S_i^a S_j^a+S_i^b S_j^b+S_i^c S_j^c) + \frac{2K}{3} \sum_{\langle ij \rangle} [S_i^c S_j^c + 2S_i^a S_j^a - \sqrt{2} (S_i^c S_j^a + S_i^a S_j^c)] + D \sum_{i} (S_i^c)^2.
\label{eq:KH_Ham_abc}
\end{align}

\section{Ferromagnetic and N\'eel instabilities in the large negative $D$ limit}

As discussed in the main text, the FM phase expands anomalously into the AFM $J$ parameter space at negative $D$ region. It is due to the reduction of spin-$1$ degrees of freedom to the Ising ones. Thus, the state of each site can take only either \eqref{eq:Isinga} or \eqref{eq:Isingb}. When $K<0$, by forming an FM order, the per-site energy of the system can be lowered by an amount $|K|$. On the other hand, when an AFM $J$ is introduced to this FM state, the per-site energy is increased by an amount $3J/2$. Thus, the instability of FM phase in the large negative $D$ limit is given by $|K|=3J/2$. This corresponds to $\theta/\pi \approx 1.81\pi$. Similarly, the instability of N\'eel phase in the FM $J$ region is estimated as $\theta/\pi \approx 0.81\pi$.

\newpage

\section{Crossover from the Kitaev spin liquid to the zero state}

\begin{figure}[tbh]
	\centering
	\includegraphics[width=0.7\columnwidth]{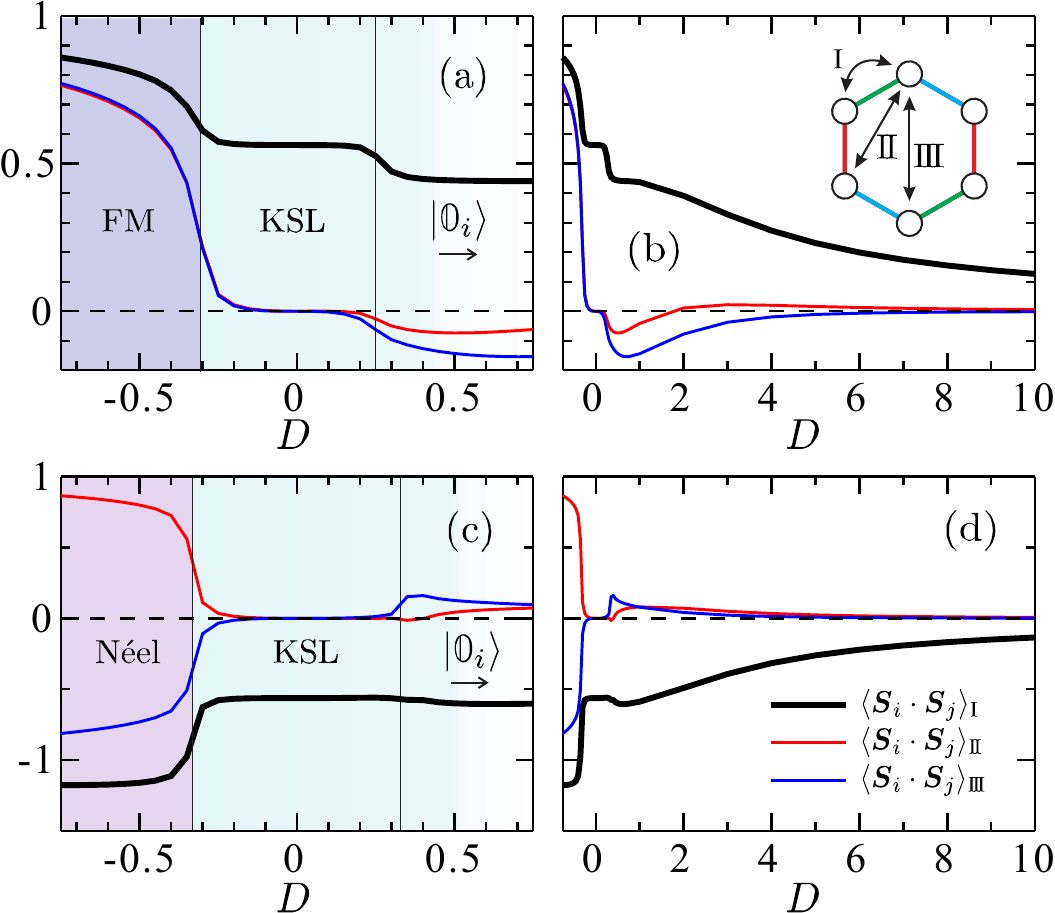} 
	\caption{Spin-spin correlation functions for nearest-neighbor, next-nearest-neighbor, and third-nearest-neighbor bonds denoted as $\langle \boldsymbol{S}_i \cdot \boldsymbol{S}_j \rangle_{\rm I}$, $\langle \boldsymbol{S}_i \cdot \boldsymbol{S}_j \rangle_{\rm I\hspace{-1.2pt}I}$, and $\langle \boldsymbol{S}_i \cdot \boldsymbol{S}_j \rangle_{\rm I\hspace{-1.2pt}I\hspace{-1.2pt}I}$, respectively. The single-ion anisotropy $D$ is varied with fixed (a,b) $\theta/\pi=1.5$ and (c,d) $\theta/\pi=0.5$.}
	\label{fig:SS}
\end{figure}

As demonstrated in the main text, starting from the Kitaev spin liquid (KSL) phase and increasing the negative single-ion anisotropy (SIA), at a certain point, the flux expectation value approaches zero abruptly, almost first-order like transition to a magnetically ordered state. Here, the strength of SIA is controlled by $D$. Conversely, with the introduction of positive SIA, the flux expectation value decreases from near $1$ to about $0.6$, after which further strengthening of SIA leads to a gradual decrease of the flux expectation value. This suggests a crossover-like transition from the KSL phase to a 'zero state' where all spin correlations vanish. To further investigate this phenomenon, we study the dependence of the spin-spin correlation functions on SIA.

Fig.~\ref{fig:SS}(a,b) plots the trend of the spin-spin correlation functions for adjacent, next-nearest, and third-nearest neighbor sites as a function of $D$ in the FM Kitaev limit ($\theta/\pi=1.5$). Fig.~\ref{fig:SS}(c,d) are similar plots in the AFM Kitaev limit ($\theta/\pi=0.5$). The average values of the spin-spin correlations are taken for the central hexagon of a 24-site lattice with open boundary conditions (see main text). One characteristic of the KSL is that the spin-spin correlation functions are finite only between nearest-neighbor lattice sites, with all longer-range correlations falling to zero. As can be seen in Fig.~\ref{fig:SS}, at $D=0$, the values of $\langle \boldsymbol{S}_i \cdot \boldsymbol{S}_j \rangle_{\rm I}$ equal to $\pm0.562921949$, while $\langle \boldsymbol{S}_i \cdot \boldsymbol{S}_j \rangle_{\rm I\hspace{-1.2pt}I}$ and $\langle \boldsymbol{S}_i \cdot \boldsymbol{S}_j \rangle_{\rm I\hspace{-1.2pt}I\hspace{-1.2pt}I}$ are zero, and these values are nearly maintained within the KSL phase. It is observed that when $D$ becomes negative, there is a rapid change in these correlations as the system transitions to FM or N\'eel phases. On the other hand, the transition on the positive side of $D$ is not as pronounced. The correlations between next-nearest neighbors and third-nearest neighbors remain small, and the nearest-neighbor correlations slowly approach zero. Considering the flux expectation values also gently converge towards zero, as demonstrated in the main text, it may be suggested that the crossover range between the KSL phase and the disordered zero-state phase is quite extensive.

\newpage

\section{Comparison with Prior Research at $D=0$}

\begin{figure}[tbh]
	\centering
	\includegraphics[width=0.7\columnwidth]{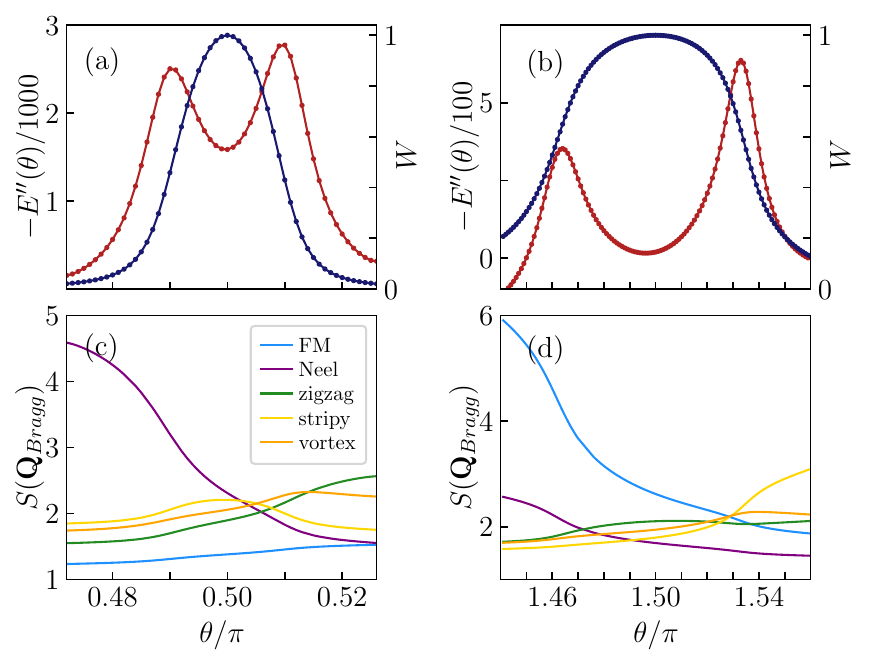} 
	\caption{DMRG results for the second derivative of ground-state energy, expectation value of the flux operator (top), intensity of the spin structure factor at representative Bragg peak positions (bottom) as a function of $\theta$ at $D=0$ around (a,c) AFM and (b,d) FM Kitaev points.}
 
	\label{fig:D0}
\end{figure}

We here evaluate our computational model using a 24-site cluster with $C_3$ symmetry and open boundary conditions as a benchmark. We compare our calculated ground-state phase diagrams at zero single-ion anisotropy ($D=0$) with those obtained from previous work using the infinite density-matrix renormalization group (iDMRG) method~\cite{dong2020spin} and the pseudofermion functional renormalization group (PFFRG) method~\cite{fukui2022ground}. In Fig~\ref{fig:D0}, we plot the second derivative of the ground-state energy $E^{\prime\prime}(\theta)$, the expectation value of the flux operator $\langle W_{\rm p} \rangle$, and the intensity of the static spin structure factor at representative Bragg peak positions are plotted as a function of $\theta$ around the FM and AFM Kitaev points.

Our analysis indicates that the Kitaev spin liquid (KSL) phases are stable within the intervals [1.464$\pi$:1.533$\pi$] and [0.490$\pi$:0.510$\pi$] for the FM and AFM Kitaev points, respectively. These intervals are positioned between the critical values reported by the iDMRG method ([0.494$\pi$:0.506$\pi$] and [1.485$\pi$:1.514$\pi$]) and those by the PFFRG method ([0.474$\pi$:0.527$\pi$] and [1.432$\pi$:1.556$\pi$]).

The iDMRG study, which used infinite-length but narrow cylinders, tends to favor commensurate magnetic ordering over spin liquid states, potentially leading to an underestimation of the extent of the KSL phases. Conversely, the PFFRG method may overestimate the range of KSL phases, as evidenced by its application to the spin-1/2 Kitaev-Heisenberg model. Based on these comparisons, we propose that our calculations offer a more accurate estimation of the critical points for KSL phases, despite the limitations imposed by the finite size of our cluster.

\newpage

\section{Finite-size scaling of the ground-state energy at the Kitaev point}

\begin{figure}[tbh]
	\centering
	\includegraphics[width=0.45\columnwidth]{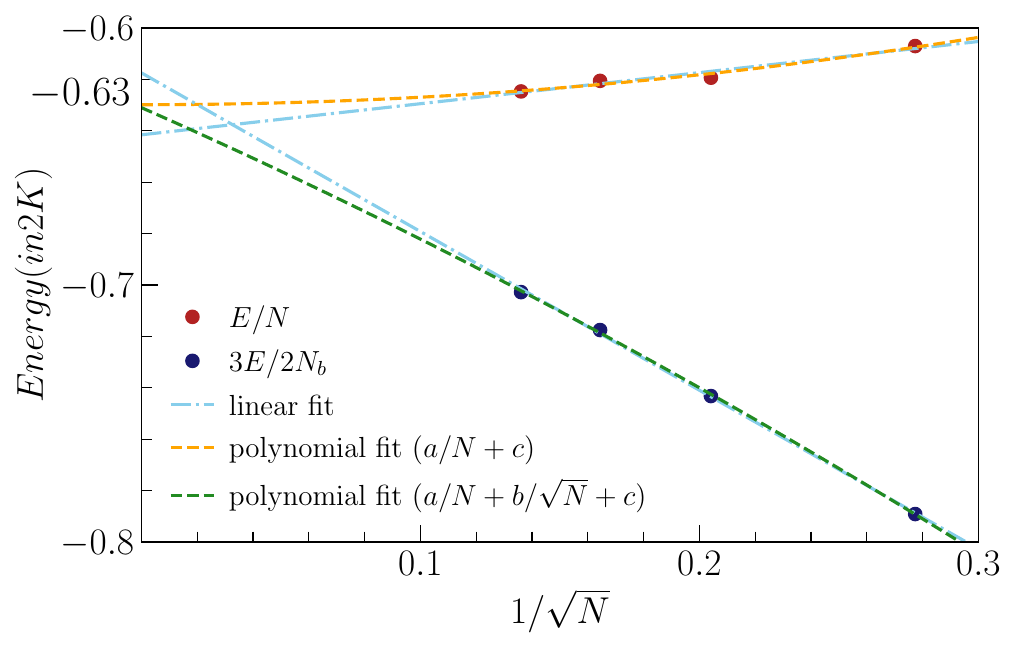} 
    \hspace{1cm}
    \includegraphics[scale= 0.3, height=0.35\columnwidth]{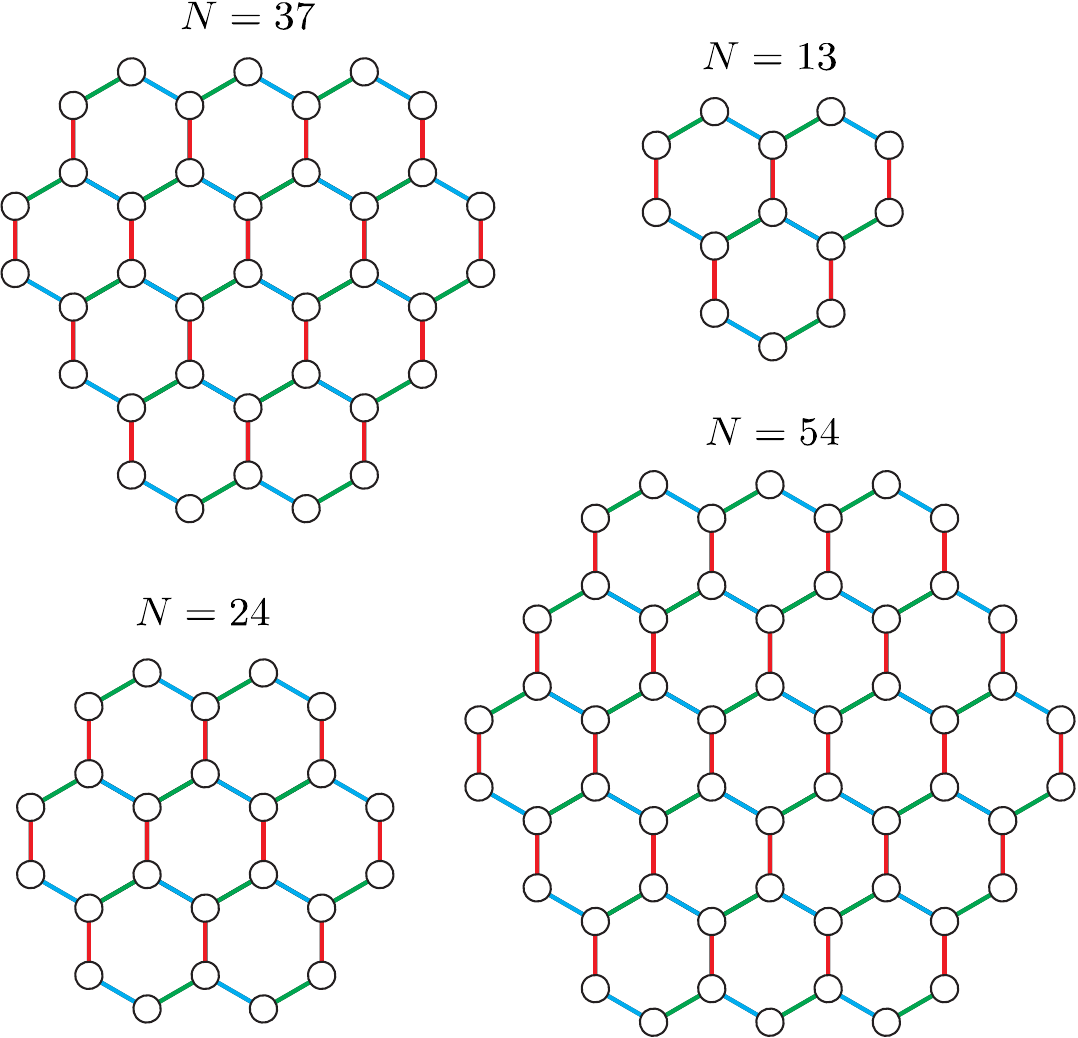}
    \caption{(left) DMRG results for energy per spin and energy per effective spins at the Kitaev point $(\theta =1.5 \pi)$ for various system sizes $N=13,24,37$ and $54$. (right) Lattice structures for $N=13,24,37$ and $54$ OBC clusters.}
    \label{fig:ener_sc}
\end{figure}

It is useful to obtain the ground-state energy at the Kitaev point. From the energy value, we can estimate the nearest-neighbor spin-spin correlations. Note that we have confirmed that longer-range spin-spin correlations are zero. In Fig.~\ref{fig:ener_sc}, we perform the finite-size scaling analysis of the ground-state energy per site. We calculate the energy per site for a finite-size cluster with open boundary conditions (OBC) in two ways: one is simply $E/N$, and the other is $\frac{3E}{2N_b}$, where $N_b$ is the number of bonds. They are extrapolated to the same value in the thermodynamic limit. We use various OBC clusters with system sizes $N=13, 24, 37$, and $54$ (see Fig. \ref{fig:ener_sc}). Since $N=13$ and $37$ clusters are $C_3$ symmetric while N=24 and 54 are $C_6$ symmetric, very precise scaling may be not expected. Nevertheless, a reasonable scaling analysis can be performed for both $E/N$ and $\frac{3E}{2N_b}$ as shown in Fig.~\ref{fig:ener_sc}. We obtain $E/N=-0.63\pm0.01$ in the thermodynamic limit. This extrapolated value is somewhat higher than the one obtained using exact diagonalization for clusters with PBC, $-0.65$ ~\cite{koga2018ground}. This may imply that the ground-state energy per site is increased by increasing system size. Base on the energy value, we estimate the nearest-neighbor spin-spin correlations as $|\langle S^\gamma_i S^\gamma_j \rangle|=0.42\pm0.003$ at the Kitaev point.

\newpage

\section{Convergence of structure factor with bond dimension $\chi$}

\begin{figure}[tbh]
	\centering
	\includegraphics[width=0.6\columnwidth]{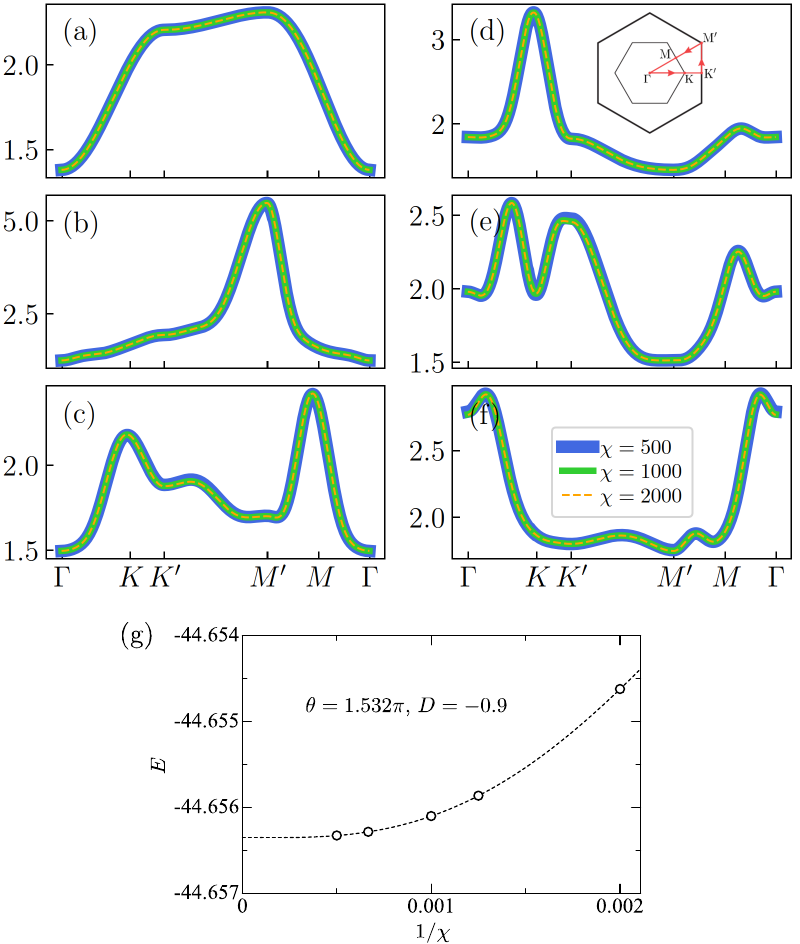} 
    \hspace{1cm}
    \caption{Variation of spin structure factor with bond dimension $\chi$ for magnetic phases (a) KSL $(\theta=0.5 \pi, D=0.0)$ (b) N\'eel $(\theta=0.48 \pi, D=0.5)$  (c) zigzag $(\theta=0.52 \pi, D=-0.5)$  (d) vortex $(\theta=1.57 \pi, D=0.5)$  (e) incommensurate $(\theta=1.560 \pi, D=-0.9)$  (f) spin liquid $(\theta=1.532 \pi, D=-0.9)$. (g) $\chi$-scaling of the ground-state energy for  $\theta=1.532\pi$ and $D=-0.9$. In the inset of (b) we show the momentum path in the Brillouin zone. }
    \label{fig:chi_scaling}
\end{figure}

In our study, while obtaining most results with the 24-site OBC cluster as presented in the main text, the bond dimension was set to 
$\chi=1000$. To examine the variation of the spin structure factor with $\chi$, we plot the evolution of the structure factor for various magnetic phases with $\chi=500$, $1000$, and $2000$ in Fig.~\ref{fig:chi_scaling}(a-f). For these values of $\chi$, we see no significant differences between different $\chi$. Consequently, it may be sufficient to use $\chi=500$ specifically for examining the spin structure factor. However, regarding the values of energy that we used to estimate the phase boundaries, there might be a larger dependence on $\chi$. Therefore, we plot the values of total energy for $\theta=1.532\pi$ and $D=-0.9$ as a function of $1/\chi$ in Fig.~\ref{fig:chi_scaling}(g). This reveals a rapid convergence of energy between $\chi=500$ and $1000$. The energy for $\chi=1000$ is $E=-44.656101602379$ and the extrapolated value to the $\chi \to \infty$ is $E=-44.6563512$, leading to an error $\Delta E=0.000249597621$. This error is sufficiently smaller than the energy differences between the value for $\theta=1.532\pi$ and those for the neighboring parameters: $E=-44.901142875433$ ($\theta=1.528\pi$) and $E=-44.424004096183$ ($\theta=1.536\pi$) as well as the difference of neighboring $\theta$ values $\Delta \theta=0.004$ when the numerical differential is performed. Thus, it would be reasonable to use this cutoff $\chi=1000$ to compute the second derivative of energy for obtaining the phase diagrams discussed in the main text.

\newpage

\section{System-size dependence of the spin structure factor for each phase}

\begin{figure}[H]
	\centering
	\includegraphics[width=0.7\columnwidth]{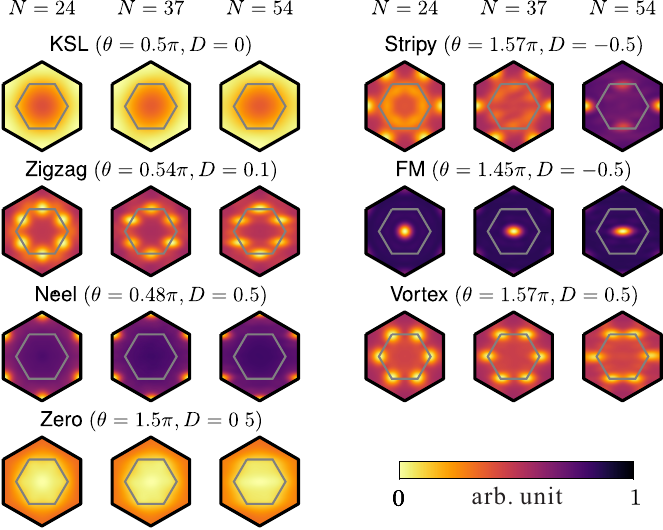} 
	\caption{Spin structure factor for various phases with varying system sizes N=24,37 and 54. The intensity of structure factor is normalized by its maximum value for each case.}
	\label{fig:ss}
\end{figure}

In the main text, we present comprehensive ground-state phase diagrams based on 24-site OBC clusters. To validate the robustness of our analysis with this cluster size, we compare the spin structure factors for system sizes $N=24, 37$, and $54$ at a representative parameter for each phase, as illustrated in Fig.~\ref{fig:ss}. The peaks of the structure factor are  qualitatively consistent across all sizes. We note that the spatial-rotational symmetry is broken in the case of stripy state with $N=54$. It may be frequently observed in DMRG calculations, particularly for larger systems. However, this symmetry breaking also serves as evidence supporting the robustness of an ordered state.

\newpage

\section{How to distinguish between vortex-I and vortex-II states}

\begin{figure}[H]
	\centering
	\includegraphics[width=0.6\columnwidth]{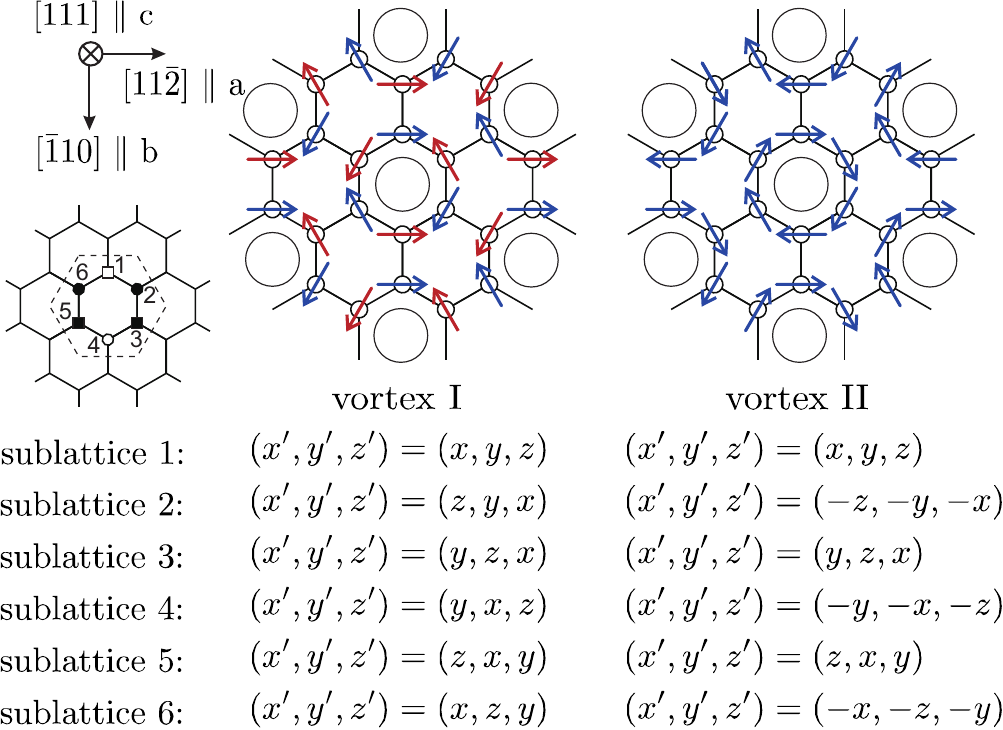} 
	\caption{Spin structures of vortex-I and vortex-II states, and the corresponding six-sublattice transformation ${\cal T}_6$.
 }
	\label{fig:vortex}
\end{figure}

Given that the spin structure factors for vortex-I and vortex-II states possess identical Bragg peak positions, further ingenuity is required to differentiate these vortex states. We here achieve it by employing a pinning-spin technique. The six-sublattice transformation ${\cal T}_6$ for the vortex states is shown in Fig.~\ref{fig:vortex}~\cite{Chaloupka2015}. Here, we specifically pin the spin of a single site in sublattice 1 along the [$11\bar{2}$] direction. This strategic approach allows for a direct comparison of our spin structure with the six-sublattice transformation ${\cal T}_6$. When the parameters are set to $\theta=1.55$ and $D=0.7$, we determine the spin components for each sublattice as follows:
\begin{align}
\nonumber
&{\rm sublattice 1:}\: (x',y',z')=(0.4087,0.408,-0.8160)\\\nonumber
&{\rm sublattice 2:}\: (x',y',z')=(-0.6616,0.4494,0.2745)\\\nonumber
&{\rm sublattice 3:}\: (x',y',z')=(0.3314,-0.5794,0.2718)\\\nonumber
&{\rm sublattice 4:}\: (x',y',z')=(0.3231,0.3231,-0.5617)\\\nonumber
&{\rm sublattice 5:}\: (x',y',z')=(-0.5794,0.3314,0.2718)\\\nonumber
&{\rm sublattice 6:}\: (x',y',z')=(0.4494,-0.6616,0.2745),
\end{align}
which correspond to the vortex-I state. Similarly, for $\theta=0.52$ and $D=0.7$, the spin components are obtained as:
\begin{align}
\nonumber
&{\rm sublattice 1:}\: (x',y',z')=(0.4089,0.4089,-0.8158)\\\nonumber
&{\rm sublattice 2:}\: (x',y',z')=(0.6315,-0.4780,-0.1600)\\\nonumber
&{\rm sublattice 3:}\: (x',y',z')=(0.3565,-0.5729,0.1704)\\\nonumber
&{\rm sublattice 4:}\: (x',y',z')=(-0.3432,-0.3432,0.5504)\\\nonumber
&{\rm sublattice 5:}\: (x',y',z')=(-0.5729,0.3565,0.1704)\\\nonumber
&{\rm sublattice 6:}\: (x',y',z')=(-0.4780,0.6315,-0.1600)
\end{align}
corresponding to the vortex-II state.

\newpage